\documentclass[onecolumn]{elsart}
\usepackage{graphicx}
\journal{New Astronomy}

\def\bibcode#1{(\texttt{#1})}
\begin{document}

\begin{frontmatter}
\title{Density profiles of dark matter halos with anisotropic velocity
       tensors}
   \author{N. Hiotelis\thanksref{now}}
\address{1st Experimental Lyceum of Athens, Ipitou 15, Plaka, 10557, Athens,
   Greece, E-mail: hiotelis@avra.ipta.demokritos.gr}

\thanks[now]{Present address: Roikou 17-19, Neos Kosmos, Athens, 11743
Greece}
\begin{small}
\begin{abstract}
      We present density profiles, that are solutions of the spherical Jeans equation, derived under
    the following two assumptions: (i) the coarse grained phase-density follows
    a power-law of radius, $\rho/{\sigma}^3\propto r^{-\alpha}$,
and (ii) the velocity anisotropy parameter
   is given by the relation $\beta_a(r)=\beta_1+2\beta_2\frac{r/r_*}{1+(r/r_*)^2}$  where
   $\beta_1$, $\beta_2$ are  parameters and  $r_*$ equals twice the virial radius,
   $r_{vir}$, of  the system. These assumptions are well motivated by the results of
   N-body simulations. Density profiles have increasing logarithmic slopes $\gamma$,
    defined by $\gamma=-\frac{\mathrm{d}\ln\rho}{\mathrm{d}\ln r}$.
    The values of $\gamma$ at $r=10^{-2.5}r_{vir}$, a distance where the
    systems could be resolved by large N-body simulations, lie in the range $1. - 1.6$.
     These inner values of $\gamma$ increase
    for increasing $\beta_1$ and for increasing concentration of the system. On the other
    hand, slopes at $r=r_{vir}$ lie in the range $2.42 - 3.82$. A model density profile
    that fits well the results at radial distances between $10^{-3}r_{vir}$ and $r_{vir}$
     and connects kinematic and structural characteristics of
     spherical systems is described.

\end{abstract}
\end{small}

\end{frontmatter}
%
%________________________________________________________________

\section{Introduction}

Density profiles of galactic halos are well fitted  by two-power
law models. Numerical studies (Quinn et al. \cite{quinn}; Frenk et
al. \cite{frenk}; Dubinski \& Garlberg \cite{dubinski}; Crone et
al. \cite{crone}; Navarro et al. \cite{navarro}, NFW; Cole \&
Lacey \cite{cole}; Huss et al. \cite{huss}; Fukushige \& Makino
\cite{fuku}; Moore et al. \cite{moore}, MGQSL; Jing \& Suto
\cite{jing}, JS) showed that the profile of relaxed halos steepens
monotonically with radius. The logarithmic slope
$\gamma=-\frac{\mathrm{d}\ln\rho}{\mathrm{d}\ln r}$ is less than 2
near the center and larger than 2 near the virial radius of the
system. The value of $\gamma $ near the center of the halo is not
yet known. Navarro et al. \cite{navarro}  claimed $\gamma=1$
while Kravtsov et al. \cite{kravtsov} initially claimed
$\gamma\sim 0.7$ but in their revised conclusions (Klypin et al.
\cite{klypin}) they argue that the inner slope varies from 1 to
1.5. Moore et al.  \cite{moore}  found a slope $\gamma=1.5$ at the
inner regions of their N-body systems.

 N-body simulations give also valuable information about
the kinematic state of the relaxed structures. Taylor \& Navarro
\cite{taylor}, (TN) showed that in a variety of cosmologies the
velocity dispersion $\sigma^2$ and the density $\rho$ at distance
$r$ are connected by a scale free relation $\rho/\sigma^3\propto
r^{-\alpha}$ with $\alpha=1.875$. They also examined density
profiles that resulted from solving the spherical Jeans equation
under the above condition assuming an isotropic velocity
distribution. Final structures of N-body simulations are radially
anisotropic. The radial anisotropy increases outwards taking its
maximum value (that is in the range $0.3 - 0.8$) at about twice
the virial radius of the the system. This behavior seems to be
universal (Cole \& Lacey \cite{cole}, Colin et al. \cite{colin},
Carlberg et al. \cite{carlberg}).

In this paper we present solutions of spherical Jeans equation
using the above scale free relation given in TN and a model for
the velocity anisotropy parameter, in order to construct density
profiles for radially anisotropic models. The method for the
solution of Jeans equation in described in Sect. 2 and the results
are presented in Sect. 3. The conclusions are discussed in Sect.
4.

\section{The solution of Jeans equation}

Under the assumption of spherical symmetry and no rotation the
Jeans equation is:

\setcounter{equation}{0}
\begin{equation}
\frac{1}{\rho(r)}\frac{\mathrm{d}}{\mathrm{d}r}(\rho(r){\sigma}_r^2(r))
+\frac{2\beta_{a}(r){\sigma}_r^2(r)}{r}=-\frac{\mathrm{d}\Phi(r)}{\mathrm{d}r}=
-\frac{GM(r)}{r^2}, \label{eqb1}\\
\end{equation}

\noindent where  $\beta_{a}$ is the velocity anisotropy parameter
given by the relation $\beta_{a}\equiv
1-\frac{{\sigma}^2_t(r)}{2{\sigma}_r^2(r)}$ (where ${\sigma}^2_t$
and ${\sigma}^2_r$ are the  tangential and the radial dispersion
of velocities respectively) and $\Phi$ is the potential.
Multiplying both sides of (\ref{eqb1}) with $r^2$ and
differentiating with respect to $r$ we have\

\begin{equation}
\frac{\mathrm{d}}{\mathrm{d}r}\left[{\frac{r^2}{\rho(r)}\frac{\mathrm{d}}{\mathrm{d}r}(\rho(r)
{\sigma}^2_r(r)) +2r \beta_{a}(r){\sigma}^2_r(r)}\right]= -4\pi
G\rho(r)r^2.
\end{equation}
Under the assumption
\begin{equation}
{\rho(r)}/{\sigma}^3(r)=(\rho(r_0)/{\sigma}^3(r_0))(r/{r_0})^{-\alpha},
\end{equation}
and after setting $\rho_0\equiv \rho(r_0), \sigma_0\equiv
\sigma(r_0), x\equiv\ln(r/r_0)$ and $y\equiv\ln(\rho/{\rho}_0)$
Jeans equation is written in the form of the following set of
differential equations:

\begin{equation}
y'_{2}(x)=-ke^{3x+y_1(x)},
\end{equation}
with $y_1(x)\equiv y(x)$ and
\begin{equation}
y'_1(x)=\frac{3}{5}\left[[3-2\beta_{a}(x)]y_2(x)
e^{-(\frac{2\alpha+3}{3}x+\frac{2y_1}{3})}-\frac{2\alpha}{3}
-2\beta_{a}(x)-2\frac{\mathrm{d}\beta_{a}(x)}
{\mathrm{d}x}[3-2\beta_{a}(x)]^{-1}\right],\label{eqb5}\\
\end{equation}
where $k\equiv \frac{4\pi G\rho_0 r^2_0}{\sigma^2_0}$. Note that
$\sigma$ in (3)
is the total dispersion of velocities.\\
TN showed that the isotropic case ($\beta_{a}\equiv 0$) admits a
solution $y_1(x)=-\beta x$, (power law in $r-\rho$ space), with
$\beta=6-2\alpha$ and $k=2(\beta-1)(3-\beta)/3$. Additionally they
used $\alpha=1.875$, a value resulting from the final structures
of N-body simulations, to show that for values of $k$ larger than
the one that corresponds to the above power-law solution, the
density profiles become complex and that for values of $k$ larger
than a critical value $k_{crit}$, the density profiles become
unrealistic since densities are vanished at some finite radius
near the center of the system. The density profile that
corresponds to $k_{crit}$ scales as $r^{-0.75}$ in the inner
region of the system, close to the NFW profile that scales as
$r^{-1}$.

Since the anisotropy of velocities is always present in the
results of N-body simulations, it should be interesting to study
its effects on the solutions of Jeans equation. N-body simulations
for a variety of cosmologies show an almost universal variation of
$\beta_{a}$. Starting from the center of the system, $\beta_{a}$
increases outwards taking its maximum radius at about twice the
virial radius $r_{vir}$ of the system. The maximum value is in the
range 0.3-0.8 (Cole \& Lacey \cite{cole}, Colin et al.
\cite{colin}, Carlberg et al. \cite{carlberg}). Thus we model
$\beta_{a}$ as
\begin{equation}
\beta_a(r)=\beta_1+2\beta_2\frac{r/r_*}{1+(r/r_*)^2}
\end{equation}
where $r$ is the radial distance. The maximum value of
$\beta_{a}$ is $\beta_{max}=\beta_1+\beta_2$ and it is
achieved at $r=r_*=2r_{vir}$. The initial conditions required for
the solution of the above set of differential equations are set in
a similar way as in TN: At a radius $r_0=1$ we assume a density
$\rho_0 = 1$ and $\gamma(r_0=1)= 2.25$. Thus we take $y_1(0)=0$
and $y'_1(0)=-2.25$. Furthermore, the value of $y_2(0)$ is
calculated by solving equation (\ref{eqb5}) for $y_2$ leading to
the relation:
\begin{equation}
y_2(0)=\frac{1}{3-\beta_a(0)}\left[\frac{2\alpha}{3}-\frac{5\beta}{3}
+2\beta'_{a}(0)[3-2\beta_a(0)]^{-1}+ 2\beta_a(0)\right]
\end{equation}
Obviously $y'_2(0)=-k$.  Then we integrate inwards and outwards to
calculate density profiles.
%__________________________________________________________________

\begin{figure}[b]
\includegraphics[width=14cm]{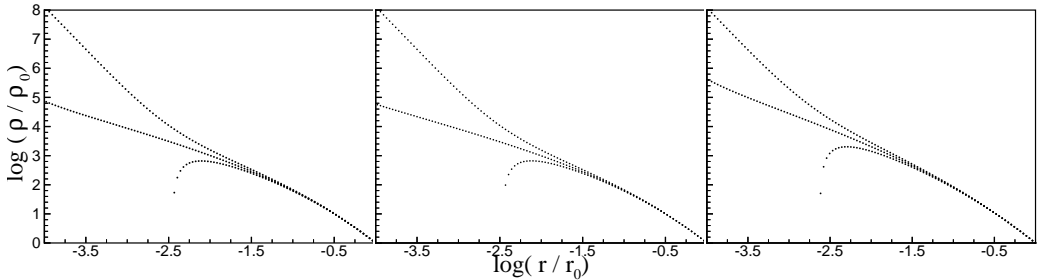}
\caption{Density profiles derived from the solution of Jeans
equation for fixed values of $\beta_1, \beta_2$ and $c$ and
varying $k$. For values of $k$ larger than a $k_{crit}$ densities
vanish at a finite radius near the center while for $k< k_{crit}$
the slopes increase inwards.} \label{fig1}
\end{figure}

\section{Density profiles}
The calculation of density profiles is organized as follows: For
given values of $r_{vir},  \beta_1, \beta_2$ we find the  solution
of Jeans equation for the maximum value of $k$, named $k_{crit}$,
for which the density is a decreasing function of the radial
distance. An example of the behavior of the solutions in the inner
region of the system is presented in Fig. \ref{fig1}. The curves
shown in the left panel are solutions for $\beta_1=0, \beta_2 =0.8
$ and $r_{vir}/r_0=10$. The upper line corresponds to $k=0.84$,
the middle one is the critical solution corresponding to
$k=k_{crit}=0.85114$ and the lower one corresponds to $k=0.86$.
The middle panel shows solutions for $r_{vir}/r_0=5, \beta_1 =0 $
and $\beta_2=0.3$. The upper line corresponds to $k=0.85$, the
middle one is the critical solution corresponding to $k=0.86146$
and the lower one to $k=0.87$. The right panel contains the
solutions for $\beta_1=0.15, \beta_2=0.35$ and $r_{vir}/r_0=10$.
The critical solution (middle curve) is obtained for
$k=k_{crit}=0.80031$ while the upper and lower curves are
solutions for $k=0.79$ and $k=0.81$ respectively. It is shown in
this Fig. that for values of $k$ smaller than $k_{crit}$ the slope
of the density profile changes significantly at the inner region
of the system where it becomes very steep, while for $k >
k_{crit}$ density profile becomes unrealistic since it vanishes at
a finite distance from the center. Although there is no obvious
physical reason to exclude solutions with $k< k_{crit}$ we note
that such a behavior of the density profile has not been reported
elsewhere and so we restrict ourselves to the study of critical
solutions alone.

We studied  $21$ different cases. The values of the parameters for
every case as well as the resulting values of $k_{crit}$ are given
in the Table \ref{Tab1}. A convenient  concentration parameter $c$
is defined by the relation $c\equiv r_{vir}/r_0$, that is the
ratio of the virial radius to the radius where the logarithmic
slope of the density profile equals to $2.25$.

\begin{figure}[b]
\includegraphics[width=14cm]{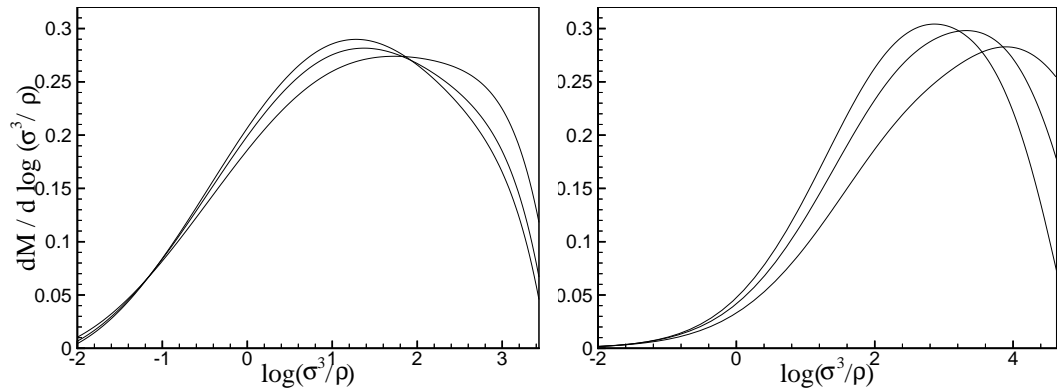} \caption{The phase-space
distribution function for anisotropic cases.  The most peaked
curves correspond to the critical values of $k$ for every case.The
left panel presents systems with $c=10$ $b_1=0$ and $b_2=0.8$
while the right one shows systems with $c=10$ $b_1=0.15$ and
$b_2=0.35$. The systems of every panel have the same mass and
energy. }\label{fig2}
\end{figure}

TN illustrated that the critical solution ($k=k_{crit}$) for the
isotropic velocity case yielded a phase-space distribution
function ($\mathrm{d}M/\mathrm{d }log({\sigma}^3/\rho)$) that was
the most peaked, or 'maximally mixed' (excluding cases that
produced hollow-core density profiles). We show that the same
holds in anisotropic cases. The left panel of Fig. \ref{fig2}
presents the phase space distribution function for the case with
$c=10$, $b_1=0$ and $b_2=0.8$. The most peaked curve corresponds
to $k=k_{crit}=0.85114$ the middle one to $k=0.84$ and the lower
one to $k=0.82$. The right panel presents the case with $c=10$,
$b_1=0.15$  and $b_2=0.35$. The most peaked curve of the this
panel corresponds to $k=k_{crit}=0.80031$, the middle one to
$k=0.78$, while the lower one to $k=0.76$.

\begin{table}[t]
\caption[] {The parameters and the resulting values of $k_{crit}$
for Cases 1-21}
\label{Tab1}
\begin{center}  \begin{tabular}{ccccc|ccccc}
\hline Case No  & $c$    & $\beta_1$  & $\beta_2$  & $k_{crit}$ &
Case No  & $c$ & $\beta_1$ & $\beta_2$ & $k_{crit}$  \\ \hline
Case 1 & 10. & 0.00&0.80&0.85114   & Case 10 & 10. & 0.15 & 0.65 & 0.78410\\
Case 2 & 5.  & 0.00&0.80&0.80947   & Case 11 & 5.  & 0.15 & 0.65 & 0.74800\\
Case 3 & 2.5 & 0.00&0.80&0.72034   & Case 12 & 2.5 & 0.15 & 0.65 & 0.67110\\
Case 4 & 10. & 0.00&0.50&0.86638   & Case 13 & 10. & 0.15 & 0.35 & 0.80031\\
Case 5 & 5.  & 0.00&0.50&0.84107   & Case 14 & 5.  & 0.15 & 0.35 & 0.78150\\
Case 6 & 2.5 & 0.00&0.50&0.78992   & Case 15 & 2.5 & 0.15 & 0.35 & 0.74380 \\
Case 7 & 10. & 0.00&0.30&0.87642   & Case 16 & 10. & 0.15 & 0.15 & 0.81100\\
Case 8 & 5.  & 0.00&0.30&0.86146   & Case 17 &  5. & 0.15 & 0.15 & 0.80310\\
Case 9 &2.5  & 0.00&0.30&0.83184   & Case 18 & 2.5 & 0.15 & 0.15 & 0.78750\\
       &     &     &    &          & Case 19 &  10.& 0.30 & 0.50 & 0.71704\\
  &     &     &      &             & Case 20 &  5. & 0.30 & 0.50 & 0.68700\\
&     &     &      &               & Case 21 &  2.5& 0.30 & 0.50 & 0.62330\\
\hline
\end{tabular}
\end{center}
\end{table}

The resulting density profiles are fitted by curves of the form
\begin{equation}
\rho_{fit}(r)=\frac{\rho_c}{(\frac{r}{r_s})^{\lambda}{(1+(\frac{r}{r_s})^{\mu})}^{\nu}}.\label{eqb8}\\
\end{equation}
Models with density cusps that have been proposed in the
literature belong to the above class of density profiles. For
$\lambda=1, \mu=1$ and $\nu=2$ Eq.(\ref{eqb8}) corresponds to the
NFW model. The model proposed by Hernquist (\cite{hern}) has
$\lambda=1, \mu=1$ and $\nu=3$. MGQSL proposed the model with
$\lambda=1.5, \mu=1.5$ and $\nu=1$ while JS estimated
$\lambda=1.5, \mu=1$ and $\nu=1.5$ for galaxy-size halos. The
fitting parameters $\rho_c$, $r_s$, $\lambda$,  $\mu$ and $\nu$ of
Eq.(\ref{eqb8}) are calculated finding the minimum of the sum
\begin{equation}
S=\sum_{i=1}^{NP}\left[\log\rho_{J}(r_i)-\log\rho_{fit}(r_i)\right]^2
\end{equation}
where $\rho_{J}$ is the density profile predicted by the above
described  solution of Jeans equation. The minimum of $S$ is found
using the unconstrained minimizing subroutine ZXMWD of IMSL
mathematical library.

\begin{figure}[b]
\includegraphics[width=14cm]{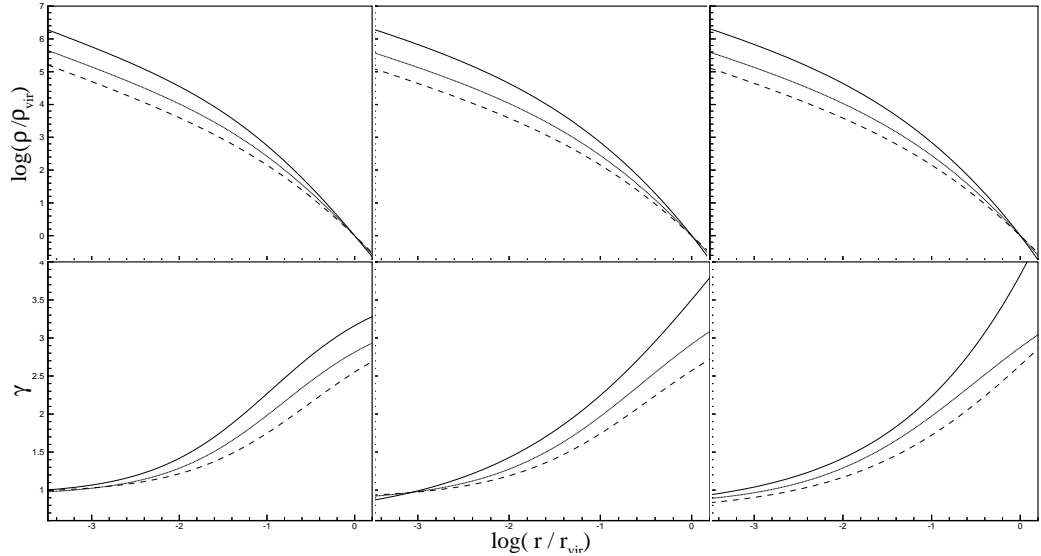} \caption{Density profiles and their logaritmic
slopes for Cases 1 to 9. First column: Cases 1, 2 and 3 (solid,
dotted and dashed line respectively). Second column: as in the
first column but for Cases 4, 5, 6. Third column: as in the first
column but for Cases 7, 8, 9.} \label{fig3}
\end{figure}

We  used values of the density at radial
distances in the range $10^{-3.5}r_{vir}$  to $1.5r_{vir}$.  The
logarithmic slope of the density profile at $r\rightarrow 0$ is
given by the value of $\lambda$. For the first nine cases, that
are all isotropic at the center ($\beta_1=0$), values of $\lambda$
that result from the above fitting are in the range $0.71 - 0.98$.
For the next nine cases (Cases 10 to 18) that have $\beta_1=0.15$
the values of $\lambda$ are in the range $0.82 - 1.02$ while for
the last three cases (Cases 19, 20 and 21), that are highly
anisotropic at the center ($\beta_1=0.3$) the values of $\lambda$
are in the range $1.06 - 1.22$. We also note that in all cases the
values of $\mu$ are in the range $0.32 - 0.61$ in contrast to all
the proposed models that give $\mu\geq 1$.

\begin{figure}[t]
\includegraphics[width=14cm]{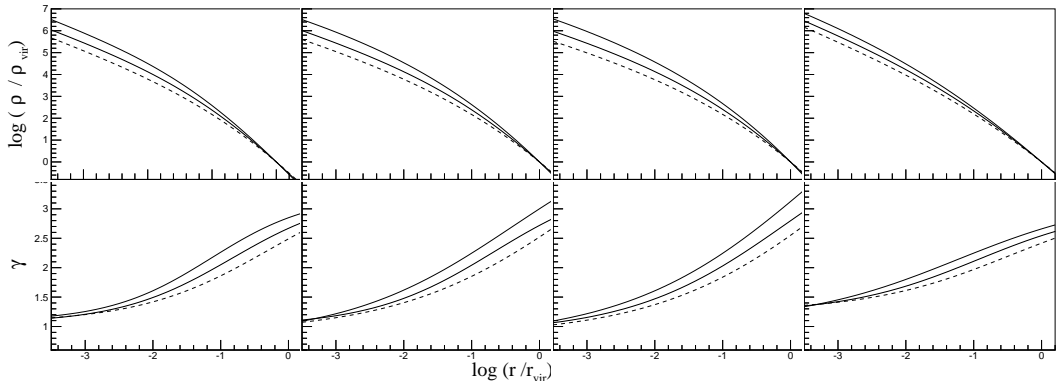}
\caption{As in Fig.\ref{fig3} but for Cases 10-21.} \label{fig4}
\end{figure}

The quality of fit is excellent and it allows a very good
estimation of the logarithmic slope of the density profile all the
way from very
small to vary large radii through the relation:\\
\begin{equation}
\gamma =-\frac{\mathrm{d}{\ln(\rho(r))}}{\mathrm{d}\ln
r}=\lambda+\mu\nu\frac{(r/r_s)^{\mu}}{1+(r/r_s)^{\mu}}
\end{equation}
The results for all  cases are presented in Fig. \ref{fig3} and
Fig. \ref{fig4}. Figure \ref{fig3} consists of six panels that
show the density profiles as well as the logarithmic slopes of
these profiles for the first nine cases. The two figures of the
first column correspond to the Cases 1, 2 and 3 (solid, dotted and
dashed lines respectively). The second column shows the results
for the Cases 4, 5 and 6 respectively while the third one contains
the results for Cases 7, 8 and 9. Densities are normalized to the
virial density of every case, that is the density of the system at
its virial radius, while distances are normalized to the virial
radius. In a similar way the rest 12 cases are presented in Fig.
\ref{fig4}. The first column shows the Cases 10, 11 and 12 (solid,
dotted and dashed line respectively). Cases 13, 14 and 15 are
shown in the two figures of the second column. The third column
shows the Cases 16, 17 and 18 and the last one corresponds to the
three last Cases 19, 20 and 21. In this way every column shows
three cases that have the same values of $\beta_1$ and $\beta_2$
but different values of $c$. Additionally, irrespectively of the
column,  solid lines correspond to systems with $c=10$, dotted
lines to systems with $c=5$ and dashed lines to systems with
$c=2.5$

The results lead to the following conclusions:
\begin{enumerate}
\item    The slope of the density profile at the central
region is closely connected to the amount of the anisotropy of
velocities there. It is  larger for larger anisotropy. For example
at $r=10^{-2.5}r_{vir}$ the first nine Cases 1 to 9 that have
$\beta_1=0.0$ show central slopes near to unity $[1-1.18]$. Cases
10 to 18 that correspond to $\beta_1=0.15$ have slopes in the
range $[1.25-1.39]$ and the last three cases, 19, 20 and 21 that
have highly anisotropic central regions, ($\beta_1=0.3$), have
slopes in the range $[1.49-1.61]$.
\item  Structures resulting for the same values of $\beta_1$
and $\beta_2$ show inner slopes that depend on their
concentrations. This dependence is not very clear at very small
radii, (for example at $r=10^{-3.5}r_{vir})$ but it is obvious at
larger radii as those that can be resolved by N-body simulations,
$(r\approx 10^{-2.5}r_{vir})$. Thus, if the values of $\beta_1$
and $\beta_2$ are universal, systems with larger $c$ have larger
slopes than those with smaller $c$ at distances that are the same
fraction of their virial radii. Note that the results of N-body
simulations suggest that $c$ is a decreasing function of the
virial mass of the system (NFW, Bullock et al.  \cite{bullock}).
Thus, our results indicate that inner slopes are smaller for
systems with larger masses. This dependence is in accordance with
the results of N-body simulations. For example JS found that, at
$r=0.01r_{vir}$, the slope is $\approx 1.5$ for galaxy-size halos
and $\approx 1.1$ for galaxy cluster-size halos respectively. We
note at this point that our concentration parameter $c$ is related
to the concentration parameter of NFW, $c_{NFW}\equiv
r_{vir}/r_{s,NFW}$, by the relation $c_{NFW}=(3/5)c$. The results
of N-body simulations indicate that for galaxy-size halos
$c_{NFW}\approx17$ (that corresponds to $c=10$), while for galaxy
cluster-size halos $c_{NFW}\approx 7$ that corresponds to
$c\approx 4.2$.
\item Slopes at $r=r_{vir}$ show values in the range
$2.42 - 3.82$. For systems with the same values of $\beta_1$ and
$\beta_2$ outer slope is an increasing function of $c$. On the
other hand, a comparison between systems with the same values of
$c$ and $\beta_1$ shows that the ones with smaller values of
$\beta_2$ have larger outer slopes.

\end{enumerate}
\begin{figure}[b]
\includegraphics[width=14cm]{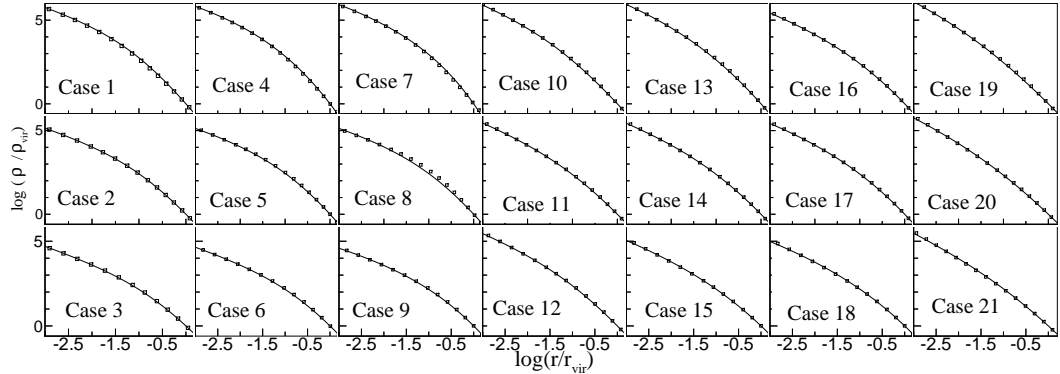}
\caption{A comparison of all resulting density profiles with our
model. Solid lines are the solutions of Jeans equation while the
predictions of our fitting models are shown by squares.}
\label{fig5}
\end{figure}

We summarize the above results by proposing a fitting formula
(FF), that holds for the range $10^{-3}r_{vir}$ to $r_{vir}$ and
connects the values of the exponents $\lambda, \mu$ and $\nu$ of
Eq. (\ref{eqb8}) with the values of $\beta_1$ and $\beta_2$. That
is:
\begin{equation}
\lambda=1+2\beta_1, \mu=0.75-0.3\beta_1,
\nu=\frac{3-2\beta_1-\beta_2}{\mu}
\end{equation}
The two remaining fitting parameters $\rho_c$ and $r_s$ are found
by the minimization procedure described above. The results are
shown in Fig. \ref{fig5}. Solid lines are the solutions of Jeans
equation while squares are the respective fits given by FF. The
fit is more than satisfactory.

Additionally, the resulting concentrations are compared to the
input concentrations by the following way: Consider $r_{n}$ the
radius where the slope equals to $n$. This $r_n$ is related to the
fitting parameters by the relation:
\begin{equation}
r_{n}=\left(\frac{n-\lambda}{\lambda+\mu\nu-n}\right)^{\frac{1}{\mu}}r_s.
\end{equation}
The concentration of the system, resulting from the fitting
procedure, is $c_f=r_{vir}/r_{2.25}$. The values of $c_f$ are
close enough to the input values of the concentration $c$. For
systems with $c=10$ the values of $c_f$ are in the range of $7.8 -
12.58$, for systems with $c=5$ the values of $c_f$ lie in the
range $4.28 - 5.33$, while the density profiles of systems with
$c=2.5$ are fitted by density profiles with $c_f$ in the range
$2.02 - 2.8$.

\begin{figure}[b]
\includegraphics[width=14cm]{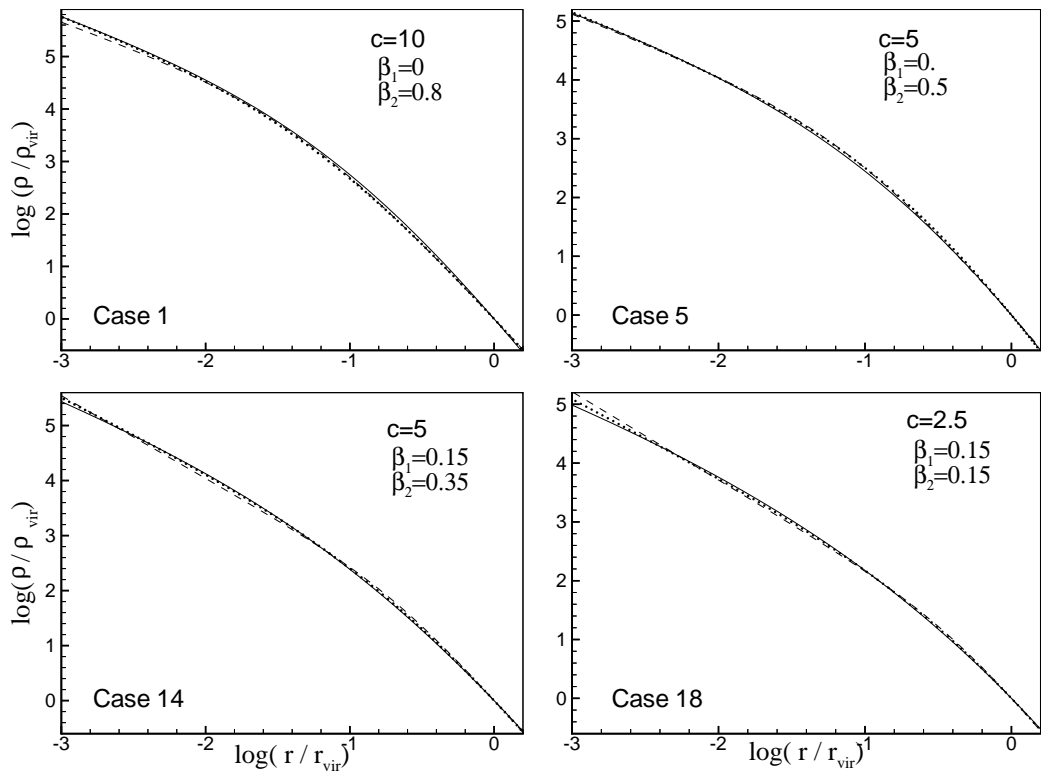}
\caption{Comparison of resulting density profiles with other
models. Solid lines: solutions of Jeans equation, dashed lines:
NFW fit (for cases 1 and 5) and MGQSL fit (for cases 14 and 18),
circles: FF fit.} \label{fig6}
\end{figure}

\begin{figure}[t]
\includegraphics[width=14cm]{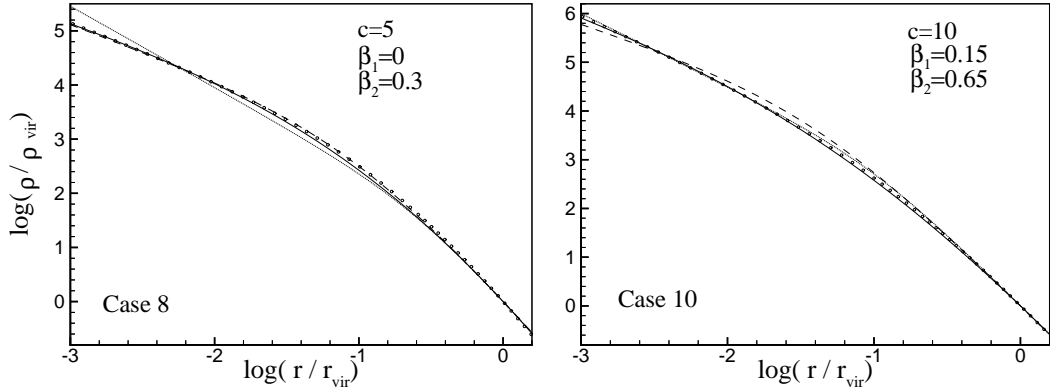}
\caption{Comparison of resulting density profiles with other
models. Solid lines: solutions of Jeans equation, dashed lines:
NFW fit, dotted lines: MGQSL fit, circles: FF fit.} \label{fig7}
\end{figure}

In Fig. \ref{fig6} a  detailed comparison between various models
is shown. For Cases 1 and 5 the solutions of Jeans equation are
shown as solid lines, the results of  FF  as dotted line and the
NFW profile as dashed lines. For cases 14 and 18 solids and dotted
lines show the same profiles as in Cases 1 and 5 while dashed
lines correspond to the MGQSL profile. It is clear that the NFW
profile is an excellent fit for the Case 5 and a very good fit for
the Case 1. On the other hand MGQSL is an excellent fit to Case 14
and a good enough fit to Case 18. The FF gives very good fits for
all four cases.  It is characteristic that analytical models with
different values of the parameters $\lambda, \mu$ and $\nu$ can
fit equally well the same density profile. This indicates the
large degree of degeneracy in these parameters that is already
noticed by Klypin et al. \cite{klypin}. However a question naturally arises
at this point. Can both popular models of NFW and MGQSL fit
equally well the same density profile? The answer is no as it can
be obtained by looking at Fig.\ref{fig7}. Solid lines are the
solutions of Jeans equation, dashed lines are the best NFW fit
while dotted lines correspond to MGQSL fit. The results of FF are
shown as small circles. For the centrally isotropic Case 8 the NFW
profile is an excellent fit. For this Case the MGQSL profile does
not approximate well the results. On the other hand the results of
Case 10, that corresponds to a centrally anisotropic system, are
fitted well by the MGQSL profile while the NFW is less steep near
the central region and also differs at intermediate distances. The
FF is a very satisfactory fit to both cases.

\section{Discussion}
This paper presents density profiles of spherical systems derived
by the solution of spherical Jeans equation. These profiles,
derived under two assumptions described in previous sections, show
a considerable agreement with models proposed in the literature.
The results show a number of trends relating kinematic and
structural characteristics of the systems. A formula that fits
very well the results for $r \geq 10^{-3}r_{vir}$ and connects the
values of the anisotropy velocity parameter to the form of the
density profile is proposed. If all cases studied above do
represent final states of real systems then the idea of a density
profile with fixed values of $\lambda, \mu$ and $\nu$ able to fit
all halos, by just adjusting the concentration, is not favored.
However it should be interesting, large N-body simulations, that
are the most powerful methods to deal with the formation of
structures, to answer some of the questions that naturally arise
by our results. Such questions relate to the validity of our two main
assumptions, the universality of the values of $\beta_1$ and
$\beta_2$ as well as the relation between the values of these two
parameters and the mass of the system. Answering  these
questions should reduce the parameter space and help to construct
better analytical models so as to improve our understanding about
the formation of structures in the universe.

\section{Acknowledgements} I would like to thank the referee James Bullock for
providing constructive comments on this manuscript and the
\emph{Empirikion Foundation} for its support.
%\end{acknowledgements}

\end{document}